\documentclass[reprint,amsmath,amssymb,aps,]{revtex4-1}
\usepackage[english]{babel}
\usepackage[utf8]{inputenc}
\usepackage{multirow}
\usepackage[colorinlistoftodos, color=green!40, prependcaption]{todonotes}
\usepackage{amsthm}
\usepackage{mathtools}
\usepackage{physics}
\usepackage{xcolor}
\usepackage{graphicx}
\usepackage[left=16mm,right=16mm,top=25mm,columnsep=20pt]{geometry} 
\usepackage{adjustbox}
\usepackage{placeins}
\usepackage[T1]{fontenc}
\usepackage{lipsum}
\usepackage{csquotes}
\usepackage[pdftex, pdftitle={Article}, pdfauthor={Author}]{hyperref} 
\begin{document}
\title{Heavy fermion representation for twisted bilayer graphene systems}

\author{Hao Shi}
\author{Xi Dai}
\altaffiliation{daix@ust.hk}
\affiliation{Department of Physics, Hong Kong University of Science and Technology, Kowloon, Hong Kong }

\begin{abstract}
We construct a heavy fermion representation for twisted bilayer graphene (TBG) systems. Two local orbitals (per spin/valley) are analytically found, which are exactly the maximally localized zero modes of the continuum Hamiltonian near the AA-stacking center. They have similar properties to the Wannier functions found in a recent study, but also have a clear interpretation as the zeroth pseudo Landau levels (ZLL) of Dirac fermions under the uniform strain field created by twisting. The electronic states of TBG can be viewed as the hybridization between these ZLL orbitals and other itinerant states which can be obtained following the standard procedure of orthogonalized plane wave method. The ``heavy fermion'' model for TBG separates the strongly correlated components from the itinerant components and provides a solid base for the comprehensive understanding of the exotic physics in TBG.
\end{abstract}

\maketitle

\section{Introduction} 
\label{sec:1}

Magic-angle twisted bilayer graphene (MATBG) has aroused continuous interest due to their rich and exotic electronic phases \cite{caoyuan1_18n,caoyuan2_18n,exp3_19sci,19n_insulator_SC_LL,20s_QAH_hBN}. These novel states are believed to closely relate to the eight flat bands near the charge neutrality point, which was first predicted by Bistritzer and MacDonald using the continuum BM model \cite{BM1_11pnas}. Extensive efforts have been made on understanding various aspects of these fascinating systems \cite{abelian_field_12prl,chiral_limit_19prl,pseudo_LL_19prb,wkb_21prl,relaxation1_17prb,relaxation2_19nm,relaxation3_19prb,relaxation4_19prr,relaxation5_20prr,relaxation6_20sa,wannier3_18prb,topology3_19prl,topology1_19prb,topology4_21prb,topology5_21prb,Hofstadter1_11prb,Hofstadter2_12prb,Hofstadter3_13prb,Hofstadter4_19prb,Hofstadter5_19prb,Hofstadter6_20prb,Hofstadter7_21prb,HF1_20prl,HF2_20prb,HF3_20prx,HF4_20prl,HF5_21prb,HF6_21prb,HF7_22prl,HF8_21prr,HF9_22prl,superconductivity1_18prx,heavy_fermion_22prl,metal1_21prl,metal2_22prl,superconductivity1_18prx,superconductivity2_18prx,superconductivity3_18prl,superconductivity4_18prl,superconductivity5_18prl,superconductivity6_19prl,sup_heavy_arxiv,Jahn_Teller1_19prx,local1_10nl,local2_12prl,local3_12prb,18prb_charge_transfer_picture,21m2D_localized_electron,BM3_12prb,exp4_19n,topology2_19prb,wannier0_18prb,wannier1_18prx,wannier2_18prx,21prr_hybrid_Wannier,BM4_21prb}.

The precious experimental and theoretical studies on MATBG indicate that both the localized and itinerant features can be found in the moir\'{e} flat bands. On the one hand, both the STM and transport measurements show Mott-like physics, suggesting very localized nature of the flat bands \cite{exp4_19n,local1_10nl,local2_12prl,local3_12prb,21m2D_localized_electron,18prb_charge_transfer_picture}. On the other hand, the topological nature found by previous theoretical studies rules out the probability that the flat bands are purely constructed by some well separated localized orbitals. It must contain some itinerant components to form the topological bands \cite{topology1_19prb,heavy_fermion_22prl,21prr_hybrid_Wannier}. The coexistence of the both components can also be seen from the typical dispersion of the flat bands: despite the complete vanishing of the Fermi velocity at the moir\'{e} Dirac points, the flat bands acquire some prominent dispersion near the moir\'{e} $\bar{\mathbf{\Gamma}}$ point, indicating considerable mixing with the itinerant components. 

A recently proposed heavy fermion model shed light on this subtle problem \cite{heavy_fermion_22prl}. In that model, two maximally localized Wannier orbitals ($f$ orbitals) are constructed using some low energy bands while all the other orbitals ($c$ orbitals) are obtained using the $k\cdot p$ expansion around the moir\'{e} Brillouin zone (mBZ) center. The complete flatness of the $f$ bands is reminiscent of the pseudo Landau level representation for TBG \cite{pseudo_LL_19prb}.

In the present paper, inspired by Ref. \cite{heavy_fermion_22prl}, we propose a more rigorous way to derive the band structure of TBG that can be expressed in terms of both the localized and itinerant basis, from a standard method developed in the early years of density functional theory, the orthogonalized plane wave (OPW) method \cite{OPW1_40pr}. In such a method, the entire crystal space is divided into two types of area: the area close to the nucleus and the interstitial area between different nucleus. In the area near the nucleus, the crystal potential is very deep, fast varying and close to a typical central potential. Therefore, the atomic wave functions can be used as a very efficient basis set to represent the eigenstates near this area, which are called core level states in solid state physics. In contrast, the potential is shallow and slowly varying in the interstitial area. It is much efficient to express the solution of the Schr{\"o}dinger equation in this area by the plane waves with an extra condition that the core level states have to be projected out from the plane wave basis used to represent the valence bands. Such a modified plane wave basis set is called OPWs, which has been further developed to the pseudo potential methods and can be viewed as part of the foundation of modern density functional theory \cite{OPW1_40pr}. Such a separation of localized (atomic like orbitals) and itinerant basis (OPW) is also a very crucial first step towards the further in-depth studies, such as LDA+DMFT and LDA+Gutzwiller, on the strongly correlated effects in many materials. In the present study, we reconstruct the moir\'{e} sub-band structure from a brand new OPW perspective. First of all, as pointed out already in our previous paper, near the AA-stacking center the TBG Hamiltonian can be approximated as Dirac electrons moving under pseudo magnetic field caused by twisting and the corresponding eigenstates are pseudo Landau levels (PLL) under the symmetric gauge condition \cite{pseudo_LL_19prb}. Among these PLLs, the zeroth PLL (ZLL) is the most localized and can be viewed as the ``core level state'' of TBG or equivalently the ``$f$ orbitals'' discussed in Ref. \cite{heavy_fermion_22prl}. Next, we construct the OPWs by projecting out these ZLL states from the plane wave basis adopted to represent the BM model. By following the standard procedure of OPW method, we can reformulate the BM model precisely into two very different basis set, the local orbitals and itinerant bands as suggested in Ref. \cite{heavy_fermion_22prl}, but without any fitting parameters to adjust. 

The localized orbitals in our approach are exactly zero modes of the Hamiltonian around AA-stacking centres. The full Hilbert space is partitioned into the localized ZLL subspace and its orthogonal subspace, the OPW subspace. By turning on the hybridization between the two subspace, the exact BM Hamiltonian can be fully restored. This method can also be applied to TBG systems with smaller angles. Based on this new representation, we can further derive the $k\cdot p$ expansion around the Dirac points and the high order magic angles can be inferred by the vanishing of the signed Fermi velocity. We can also generalize our method to analyze the band structures of the twisted multilayer graphene (TMG) systems, where large overlaps between the two low-energy bands and the localized orbitals are also observed. Finally a new mean-field variational approach can be proposed to show the important role played by the extremely localized ZLL orbitals in the correlated insulator phases for the commensurate filling cases, where the various of symmetry breaking orders are mainly taken place in ZLLs suggesting the possible emergence of strong correlation effects in these ZLLs when they are fractionally filled or at high temperature.

This paper is organized as follows. In Sec. \ref{sec:2}, a brief review of the BM Hamiltonian is given, followed by the rough demonstration of the hybridized ZLL+OPW model. In Sec. \ref{sec:3}, the model is applied in small angle TBG systems and some TMG systems. The oscillating Fermi velcocity and magic angle series are also discussed here. In Sec. \ref{sec:4}, the variational method and numerical results for MATBG at integer fillings are shown. In Sec. \ref{sec:5}, a summary is made.

\section{Formulation of the model} 
\label{sec:2}
\newcommand{\Alpha}{\mathrm{A}}
\newcommand{\Tau}{\mathrm{T}}
\newcommand{\cdummy}{\cdot}
\newcommand{\nobracket}{}
\newcommand{\tmmathbf}[1]{\ensuremath{\boldsymbol{#1}}}
\newcommand{\tmop}[1]{\ensuremath{\operatorname{#1}}}

\subsection{The continuum model}

For small-angle TBG systems, the continuum BM Hamiltonian is widely used \cite{BM1_11pnas,BM2_07prl,BM3_12prb}. The atomic valley $\eta$ is a good quantum number, giving an emergent $U_v(1)$ symmetry. We follow the formulation in Ref. \cite{wannier1_18prx} and constrain our discussion in the valley $\eta=-1$ for simplicity. The BM Hamiltonian reads
\begin{align}
  H^{\tmop{BM}} = \left(\begin{array}{cc}
    - v_F (\tmmathbf{p} - \hbar\tmmathbf{K}_1) \cdummy \tmmathbf{\sigma} & e^{-i \Delta \tmmathbf{K} \cdot \tmmathbf{r}}
    U (\tmmathbf{r})\\
    e^{i  \Delta \tmmathbf{K} \cdummy \tmmathbf{r}} U^{\dag}
    (\tmmathbf{r}) & - v_F (\tmmathbf{p} - \hbar\tmmathbf{K}_2)
    \cdot \tmmathbf{\sigma}
  \end{array}\right),
  \label{eq:BM_Hamiltonian}
\end{align}
where $\tmmathbf{p} = - i \hbar\nabla$ is the momentum operator, $v_F$ is the bare Fermi velocity, and the Pauli matrices $\tmmathbf{\sigma} =(-\sigma_x,\sigma_y)$ are defined in the space of A, B sublattice of graphene. The moir\'{e} and atomic lattice constants are $L_{\theta} = a/[2 \sin (\theta / 2)]$ and $a=0.246 $ nm, respectively. $\tmmathbf{K}_1$ and $\tmmathbf{K}_2$ are Dirac points of layer 1 and layer 2, $\Delta \tmmathbf{K}=\tmmathbf{K}_2 -\tmmathbf{K}_1 = (0,k_{\theta})$, $k_{\theta} = 4 \pi/(3 L_{\theta})$. The tunneling from layer 2 to layer 1 is described by the moir\'{e} potential $e^{-i \Delta \tmmathbf{K} \cdot
\tmmathbf{r}} U (\tmmathbf{r})$, with
\begin{align}
\begin{split}
  U (\tmmathbf{r}) = & \left(\begin{array}{cc}
    u_0 & u_1\\
    u_1 & u_0
  \end{array}\right) e^{i\tmmathbf{q}_1 \cdummy \tmmathbf{r}} +
  \left(\begin{array}{cc}
    u_0 & u_1 \omega\\
    u_1 \omega^{-1} & u_0
  \end{array}\right) e^{i\tmmathbf{q}_2 \cdummy \tmmathbf{r}} \\
  &+\left(\begin{array}{cc}
    u_0 & u_1 \omega^{-1}\\
    u_1 \omega & u_0
  \end{array}\right) e^{i\tmmathbf{q}_3 \cdummy \tmmathbf{r}},
 \end{split}
\label{eq:Moire_potential}
\end{align}
where $\omega = \exp(i2\pi/3)$ and $\tmmathbf{q}_1 = k_{\theta} (0, 1)$, $\tmmathbf{q}_2 = k_{\theta} \left(-\sqrt{3}/2,-1/2 \right)$, $\tmmathbf{q}_3 = k_{\theta} \left(\sqrt{3}/2, - 1/2 \right)$. $u_0$ and $u_1$ denote the intra- and inter-sublattice tunneling amplitudes. Usually $u_0 < u_1$ due to lattice corrugation effects \cite{wannier1_18prx}. In this paper the above parameters are fixed as $\hbar v_F = 0.5944$ eV$\cdot$nm, $u_1 = 0.11$ eV and $u_0 = 0.8 u_1$.

The BM Hamiltonian in the two valleys can be transformed to each other through $C_{2 y}$, $C_{2 z}$ or $\mathcal{T}$ (time reversal) operations, while in each valley it has $C_{3 z}$, $C_{2 x}$ and $C_{2 z} \mathcal{T}$ symmetries. For the Hamiltonian defined in Eq. (\ref{eq:BM_Hamiltonian}), there also exists an additional particle-hole symmetry $\mathcal{P}$ that guarantees the bands to be symmetric about the charge neutrality point.

\begin{figure}[ht]
	\centering
	\includegraphics[width=1.0\linewidth]{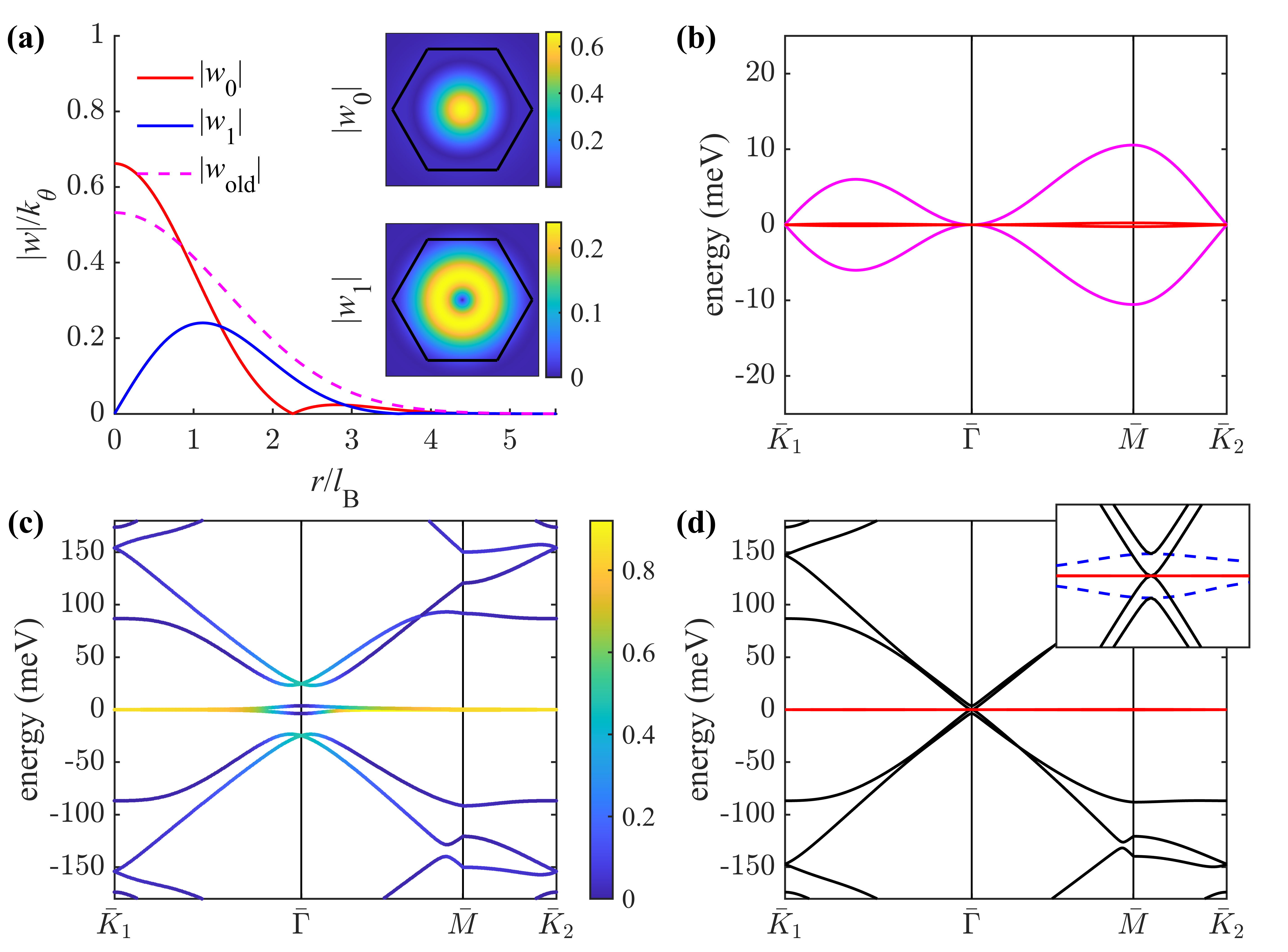}
	\caption{(a) Real space distribution of the two components $w_0(r)$ and $w_1(r)$ of ZLLs. The pink dashed line represents $w_0(r)$ if $u_0=0$ ($w_1(r)=0$ in this case). (b) The completely flat bands formed by ZLLs (\ref{eq:ZLLWF1}) and (\ref{eq:ZLLWF2}) (red) and the bands formed by the old ZLLs (with $u_0=0$) proposed in Ref. \cite{pseudo_LL_19prb} (pink). (c) The BM bands of TBG with the color representing their overlap with ZLLs: $\sum_{t}| \langle\Phi_{\bar{\tmmathbf{k}}, t} |\psi^{\tmop{BM}}_{\bar{\tmmathbf{k}}, n}\rangle|^2$. (d) The decoupled bands in the ZLL (red) and OPW subspaces (black). The inset shows the quadratic touching of the decoupled bands and the original BM bands (blue dashed lines) near the mBZ center. All figures are plotted at $\theta=1.05^\circ$.}
	\label{fg:1}
\end{figure}

\subsection{Zeroth pseudo Landau levels}

In this subsection we present the zeroth pseudo Landau level (ZLL) wave functions. Following the spirit in Ref. \cite{pseudo_LL_19prb}, first we apply the gauge transformation $\tilde{H} = V^{\dag}
(\tmmathbf{r}) H^{\tmop{BM}} V (\tmmathbf{r})$,
\begin{align}
  V (\tmmathbf{r}) = \frac{1}{\sqrt{2}} \left(\begin{array}{cc}
    e^{i\tmmathbf{K}_1 \cdummy \tmmathbf{r}} & e^{i\tmmathbf{K}_1 \cdummy
    \tmmathbf{r}}\\
    i e^{i\tmmathbf{K}_2 \cdummy \tmmathbf{r}} & - i e^{i\tmmathbf{K}_2
    \cdummy \tmmathbf{r}}
  \end{array}\right) \sigma_0,
  \label{eq:gauge_transformation}
\end{align}
which gives a Hamiltonian $\tilde{H}$ with a more symmetric form. Then we expand the moir\'{e} potential Eq. (\ref{eq:Moire_potential}) to the linear order of $\tmmathbf{r}/L_{\theta}$ around the AA-stacking center $\tmmathbf{r}=\tmmathbf{0}$. The resulting local Hamiltonian can be written as
\begin{align}
  \tilde{H}^{\tmop{AA}} = \left(\begin{array}{cc}
    - v_F \left(\tmmathbf{p} + e\tmmathbf{A}
    \right) \cdummy \tmmathbf{\sigma} & - 3 i u_0\\
    3 i u_0 & - v_F \left(\tmmathbf{p} -
    e\tmmathbf{A} \right) \cdummy \tmmathbf{\sigma}
  \end{array}\right),
  \label{eq:linear_Hamiltonian}
\end{align}
where $e$ is the elementary charge, 
$\tmmathbf{A}= B_{\theta} \left(- y/2, x/2\right)$ is the pseudo vector potential with field strength $B_{\theta} = 3 u_1
k_{\theta}/(e v_F)$. The pseudo field is locally generated by the moir\'{e} potential near the AA-stacking center, and usually it has a large magnitude. For $\theta = 1.05^{\circ}$, the field strength reaches $B_{\theta} \approx 114 $ T. 

In Ref. \cite{pseudo_LL_19prb} the intra-sublattice tunneling term $\pm 3 i u_0$ in $\tilde{H}^{\tmop{AA}}$ is dropped. Then the simplified Hamiltonian can be interpreted as two fermions coupled to the opposite magnetic fields $\tmmathbf{B}= \pm B_{\theta} \hat{\tmmathbf{e}}_z$, which has chiral zero modes that are just the zeroth Landau levels  of Dirac fermions. However, our further analysis shows that the ZLL states obtained in such an approximate way cannot be used as the efficient localized orbitals to construct OPW (see Fig. \ref{fg:1}(b)), although they are well localized around AA-stacking centers. Neglecting the intra-sublattice tunneling term is too rough an approximation for the quantitative analysis.

Fortunately, the chiral zero modes of $\tilde{H}^{\tmop{AA}}$ still exist even if the intra-sublattice tunneling is present, due to the chiral symmetry represented by the operator $\mathcal{C}=\tilde{\varrho}_z\tilde{\sigma}_z$ so that $\mathcal{C}^{-1}\tilde{H}^{\tmop{AA}}\mathcal{C}=-\tilde{H}^{\tmop{AA}}$, where $\tilde{\varrho}_z$ and $\tilde{\sigma}_z$ are Pauli matrices defined in the layer and sublattice space after the gauge transformation Eq. (\ref{eq:gauge_transformation}). After some analytical derivation and transforming back to original representation, the two maximally localized ZLL wave functions are found to be (see Supplemental Material (SM) for details)
\begin{align}
  \Phi_1 (\tmmathbf{r}) & = \frac{1}{\sqrt{2}}
  \left(\begin{array}{c}
    - e^{i\tmmathbf{K}_1 \cdot \tmmathbf{r}} e^{i \phi}
    w_1 (r)\\
    e^{i\tmmathbf{K}_1 \cdot \tmmathbf{r}} w_0 (r)\\
    - i e^{i\tmmathbf{K}_2 \cdot \tmmathbf{r}} e^{i \phi}
    w_1 (r)\\
    - i e^{i\tmmathbf{K}_2 \cdot \tmmathbf{r}} w_0 (r)
  \end{array}\right), \label{eq:ZLLWF1}\\
  \Phi_2 (\tmmathbf{r}) & = \frac{1}{\sqrt{2}} \left(\begin{array}{c}
    e^{i\tmmathbf{K}_1 \cdot \tmmathbf{r}} w_0 (r)\\
    e^{i\tmmathbf{K}_1 \cdot \tmmathbf{r}} e^{- i \phi}
    w_1 (r)\\
    i e^{i\tmmathbf{K}_2 \cdot \tmmathbf{r}} w_0 (r)\\
    - i e^{i\tmmathbf{K}_2 \cdot \tmmathbf{r}} e^{- i \phi}
    w_1 (r)
  \end{array}\right), \label{eq:ZLLWF2}
\end{align}
where $r=|\tmmathbf{r}|$, $\phi = \arg (x + i y)$. They are two generalized zeroth Landau level wave functions under the symmetric gauge with the lowest angular momentum $L_z$ and will be chosen as the two localized orbitals to construct OPWs. The functions $w_n$ ($n=0$, $1$) appearing above are
\begin{align}
w_n (r) =\mathcal{C}_0\frac{e^{\lambda_{\theta}^2 - r^2 / (4 l_B^2)}}{\sqrt{2 \pi l_B^2}} J_n \left( \frac{\sqrt{2} \lambda_{\theta} r}{l_B} \right), \label{eq:ZLLBessel}
\end{align}
where $J_n (x)$ are Bessel functions, and $\mathcal{C}_0$ is some normalization factor. We now give a further interpretation to the parameters $l_B$ and $\lambda_{\theta}$. The magnetic length $l_B = \sqrt{\hbar/(e B_{\theta})}$ quantifies the degree of localization \cite{94arxiv_qhe}. The spread of the well localized ZLLs, roughly estimated as $\sqrt{2}l_B$, is about $0.25 L_{\theta}$ for $\theta=1.05^{\circ}$ (see Fig. \ref{fg:1}(a) for the shape of $w_0(\tmmathbf{r})$ and $w_1(\tmmathbf{r})$). The dimensionless factor $\lambda_{\theta} = 3 u_0 l_B/(\sqrt{2} \hbar v_F)$ characterizes the degree of sublattice polarization. Our ZLLs will evolve to those fully sublattice-polarized ones in Ref. \cite{pseudo_LL_19prb} when $u_{0}$ is reduced to zero. More interestingly, if we expand Eqs. (\ref{eq:ZLLWF1}) and (\ref{eq:ZLLWF2}) using Landau level basis and keep only the leading terms, they reduce to the same analytical form as the (fitted) Wannier functions proposed in Ref. \cite{heavy_fermion_22prl}.

Two Bloch ZLLs are constructed by summing over all $N_{\tmop{m}}$ ZLLs located at different moir\'{e} sites $\tmmathbf{R}$, and they can finally be expressed using plane waves,
\begin{align}
\begin{split}
  | \Phi_{\bar{\tmmathbf{k}}, t} (\tmmathbf{r}) \rangle & =
  \frac{1}{\sqrt{N_{\tmop{m}}}} \sum_{\tmmathbf{R}} e^{i \bar{\tmmathbf{k}} \cdummy
  \tmmathbf{R}} | \Phi_t (\tmmathbf{r}-\tmmathbf{R}) \rangle \\
  & = \sum_{\tmmathbf{G} \alpha} \mathcal{L}_{\tmmathbf{G} \alpha, t}
  (\bar{\tmmathbf{k}}) | \bar{\tmmathbf{k}} +\tmmathbf{G}, \alpha \rangle,
\end{split}
  \label{eq:Bloch_ZLL}
\end{align}
with $t = 1$, $2$. $| \bar{\tmmathbf{k}} +\tmmathbf{G}, \alpha \rangle$ denotes the plane wave with wave vector $\bar{\tmmathbf{k}} +\tmmathbf{G}$ at layer/sublattice $\alpha$, and $\mathcal{L}
(\bar{\tmmathbf{k}})$ is the transformation matrix from plane waves to ZLLs. At each $\bar{\tmmathbf{k}}$ (we add a bar on $\tmmathbf{k}$ to indicate that it lives in the mBZ, otherwise it should be understood as a vector in the atomic BZ), the two Bloch ZLLs (\ref{eq:Bloch_ZLL}) are not strictly orthonormal, but $| \langle
\Phi_{\bar{\tmmathbf{k}}, t}| \nobracket
\Phi_{\bar{\tmmathbf{k}}, t'} \rangle \nobracket - \delta_{t t'} | \nobracket \sim 10^{- 2}$ is always satisfied for our parameters at $\theta=1.05^{\circ}$. Further normalization procedure can thus be safely neglected.

The two ZLLs account for a dominant proportion of the flat BM bands, as indicated in Fig. \ref{fg:1}(c). The overlap between the flat bands and ZLLs is relatively large near the moir\'{e} Dirac points, but approaches zero at the mBZ center, where the flat bands completely consist of itinerant states that will be introduced below. This is because the (degenerate) ZLLs form a two-dimensional representation of the group $D_3$ at $\bar{\tmmathbf{\Gamma}}$, while the (non-degenerate) BM flat bands form two one-dimensional representations there. This fact explains why some previous two-orbital tight-binding models with finite hopping cutoffs cannot fit the BM bands well near the mBZ center \cite{wannier1_18prx,wannier2_18prx,21m2D_localized_electron,18prb_charge_transfer_picture}.

We place all other PLLs in the OPW subspace introduced below since they are not well localized around the AA-stacking center. Their spatial spread is comparable with the moir\'{e} supercell and will deviate from the true eigenstates since the local Hamiltonian (\ref{eq:linear_Hamiltonian}) is reasonable only near the AA center. Incorporating them into the local orbitals will unnecessarily complicate the present model.

\subsection{Orthogonalized plane waves}

The complete Hilbert space can now be separated into two orthogonal subspaces. One of them is the ZLL subspace spanned by the localized ZLL orbitals (\ref{eq:Bloch_ZLL}). Its orthogonal complement space should contain all states that are orthogonal to the ZLLs. A general approach to construct such basis is subtracting their ZLL components from the plane waves:
\begin{align}
|\tmmathbf{k}, \alpha \rangle^{\text{opw}} \sim|\tmmathbf{k}, \alpha \rangle - \sum_{t} |\Phi_{\bar{\tmmathbf{k}}, t}\rangle \langle
\Phi_{\bar{\tmmathbf{k}}, t}\nobracket | \tmmathbf{k}, \alpha \rangle. \label{eq:opw_old}
\end{align}
Basis functions with this form was first introduced by Herring in his famous orthogonalized plane wave (OPW) method \cite{OPW1_40pr}, so we might name this subspace the OPW subspace as well.

Although mathematically clear, basis with the form Eq. (\ref{eq:opw_old}) is not convenient to use since further orthonormalization procedure is necessary. A better method is to calculate the kernel (null) space of the two Bloch ZLL wave functions (\ref{eq:Bloch_ZLL}) at each $\bar{\tmmathbf{k}}$ directly. In practical calculations, the singular value decomposition method is applied. And finally the OPW basis is formally written using plane waves as
\begin{align}
  |\Psi_{\bar{\tmmathbf{k}}, a} \rangle  =  \sum_{\tmmathbf{G} \alpha}
  \mathcal{A}_{\tmmathbf{G} \alpha, a} (\bar{\tmmathbf{k}}) |
  \bar{\tmmathbf{k}} +\tmmathbf{G}, \alpha \rangle,
  \label{eq:OPW}
\end{align}
with $a = 1$, $2$, $\ldots$, $4N_{\tmmathbf{G}}-2$, $N_{\tmmathbf{G}}$ is the number of reciprocal $\tmmathbf{G}$ vectors within the cutoff (in this work we take 61 $\tmmathbf{G}$ vectors around each atomic Dirac point). The transformation matrix $\mathcal{A}(\bar{\tmmathbf{k}})$ satisfies
\begin{align}
\begin{split}
\sum_{\tmmathbf{G} \alpha}
\mathcal{A}_{\tmmathbf{G} \alpha, a}^{\ast} (\bar{\tmmathbf{k}})
\mathcal{A}_{\tmmathbf{G} \alpha, a'} (\bar{\tmmathbf{k}}) &= \delta_{a a'},\\ 
\sum_{\tmmathbf{G} \alpha}
\mathcal{L}_{\tmmathbf{G} \alpha, t}^{\ast} (\bar{\tmmathbf{k}})
\mathcal{A}_{\tmmathbf{G} \alpha, a} (\bar{\tmmathbf{k}}) &= 0.
\end{split}
\label{eq:AA&LA}
\end{align}

\subsection{Hybridized ZLL+OPW representation}
The BM Hamiltonian (\ref{eq:BM_Hamiltonian}) is expressed under ZLL and OPW basis as (at each $\bar{\tmmathbf{k}}$)
\begin{align}
  H^{\text{BM}} (\bar{\tmmathbf{k}})  = & \sum_{t t'} \mathcal{H}^{\tmop{zll}}_{t, t'}
  (\bar{\tmmathbf{k}}) f_{\bar{\tmmathbf{k}},t}^{\dag} f_{\bar{\tmmathbf{k}},t'} + \sum_{aa'} \mathcal{H}^{\tmop{opw}}_{a, a'}
  (\bar{\tmmathbf{k}}) d_{\bar{\tmmathbf{k}},a}^{\dag} d_{\bar{\tmmathbf{k}},a'}\nonumber\\
  &  + \left( \sum_{t a} \mathcal{H}^{\tmop{cp}}_{t, a}
  (\bar{\tmmathbf{k}}) f_{\bar{\tmmathbf{k}},t}^{\dag} d_{\bar{\tmmathbf{k}},a} + h.c. \right), \label{eq:hybridized_model}
\end{align}
where $f_{\bar{\tmmathbf{k}},t}^{\dag}$ ($f_{\bar{\tmmathbf{k}},t}$) and $d_{\bar{\tmmathbf{k}},a}^{\dag}$ ($d_{\bar{\tmmathbf{k}},a}$) are creation (annihilation) operators for ZLL states and OPW states, respectively. $\mathcal{H}^{\tmop{zll}} (\bar{\tmmathbf{k}}) =\mathcal{L}^{\dag}
(\bar{\tmmathbf{k}}) \mathcal{H}^{\tmop{BM}} (\bar{\tmmathbf{k}}) \mathcal{L} (\bar{\tmmathbf{k}})\approx 0$ and $\mathcal{H}^{\tmop{opw}} (\bar{\tmmathbf{k}}) =\mathcal{A}^{\dag} (\bar{\tmmathbf{k}}) \mathcal{H}^{\tmop{BM}}
(\bar{\tmmathbf{k}}) \mathcal{A} (\bar{\tmmathbf{k}})$ are Hamiltonian kernels in the ZLL and OPW subspaces, and $\mathcal{H}^{\tmop{cp}}
=\mathcal{L}^{\dag} (\bar{\tmmathbf{k}}) \mathcal{H}^{\tmop{BM}}
(\bar{\tmmathbf{k}}) \mathcal{A} (\bar{\tmmathbf{k}})$ describes the coupling between them. Written in Eq. (\ref{eq:hybridized_model}), the BM Hamiltonian is understood as a hybridization of two localized ZLL orbitals with many itinerant OPW states.

What is really interesting is the separate band structures when we turn off the coupling between ZLLs and OPWs, i.e., when we artificially set the third term in Eq. (\ref{eq:hybridized_model}) to zero: $\mathcal{H}^{\tmop{cp}}=0$. In this case the decoupled bands in the ZLL and OPW subspaces are shown in Fig.
\ref{fg:1}(d). The well localized ZLL orbitals generate two almost completely flat bands (with maximum bandwidth $\sim 0.5$ meV), while the OPW states contribute all high-energy bands which usually have a stronger dispersion and look quite similar to the remote BM bands except near the mBZ center. The quadratic touching near the mBZ center (Fig. \ref{fg:1}(d), inset) is a salient feature, where the OPW bands behave like the energy bands of the Bernal-stacking bilayer graphene \cite{BLG_06prb} near the atomic Dirac points. It is the coupling $\mathcal{H}^{\tmop{cp}}$ that provides the exchange channel between these two subspaces, making the completely flat ZLL bands dispersive and topological, and finally splitted into the flat bands predicted by the BM model.

Before ending this section, we want to mention that the coupling term $\mathcal{H}^{\tmop{cp}}$ is the key to distinguish our model from that in Ref. \cite{heavy_fermion_22prl}. In the present study we choose to respect the actual dispersion of all high-energy bands, rather than focus only on the low-energy window near $\bar{\mathbf{\Gamma}}$ point. To recover the complete bands, many OPWs will be inevitably involved.

\section{Applications in other twisted graphene systems} 
\label{sec:3}
\subsection{Smaller angle systems}
The ZLL wave functions (\ref{eq:ZLLWF1}) and (\ref{eq:ZLLWF2}) are uniquely determined by the values of $\theta$, $v_F$, $u_0$ and $u_1$. The precise analytical form of ZLL wave functions is a huge advantage that gets us free from the Wannierization procedure in Ref. \cite{heavy_fermion_22prl} for each specific set of parameters. It is noteworthy that the ZLLs in real (reciprocal) space become even more localized (flat) in TBG with smaller twisting angles, as indicated by the magnetic length
\begin{align}
  \frac{l_B}{L_{\theta}} & = \sqrt{\frac{\hbar v_F}{4 \pi u_1 L_{\theta}}}
  \propto \frac{1}{\sqrt{L_{\theta}}} .
  \label{eq:lB/Ltheta}
\end{align}
In this subsection we extend our analysis to the second magic angle, where the Fermi velocity at the Dirac points vanishes again but the lowest two bands are no longer gapped from other bands. 

\begin{figure}[ht]
	\centering
	\includegraphics[width=1.0\linewidth]{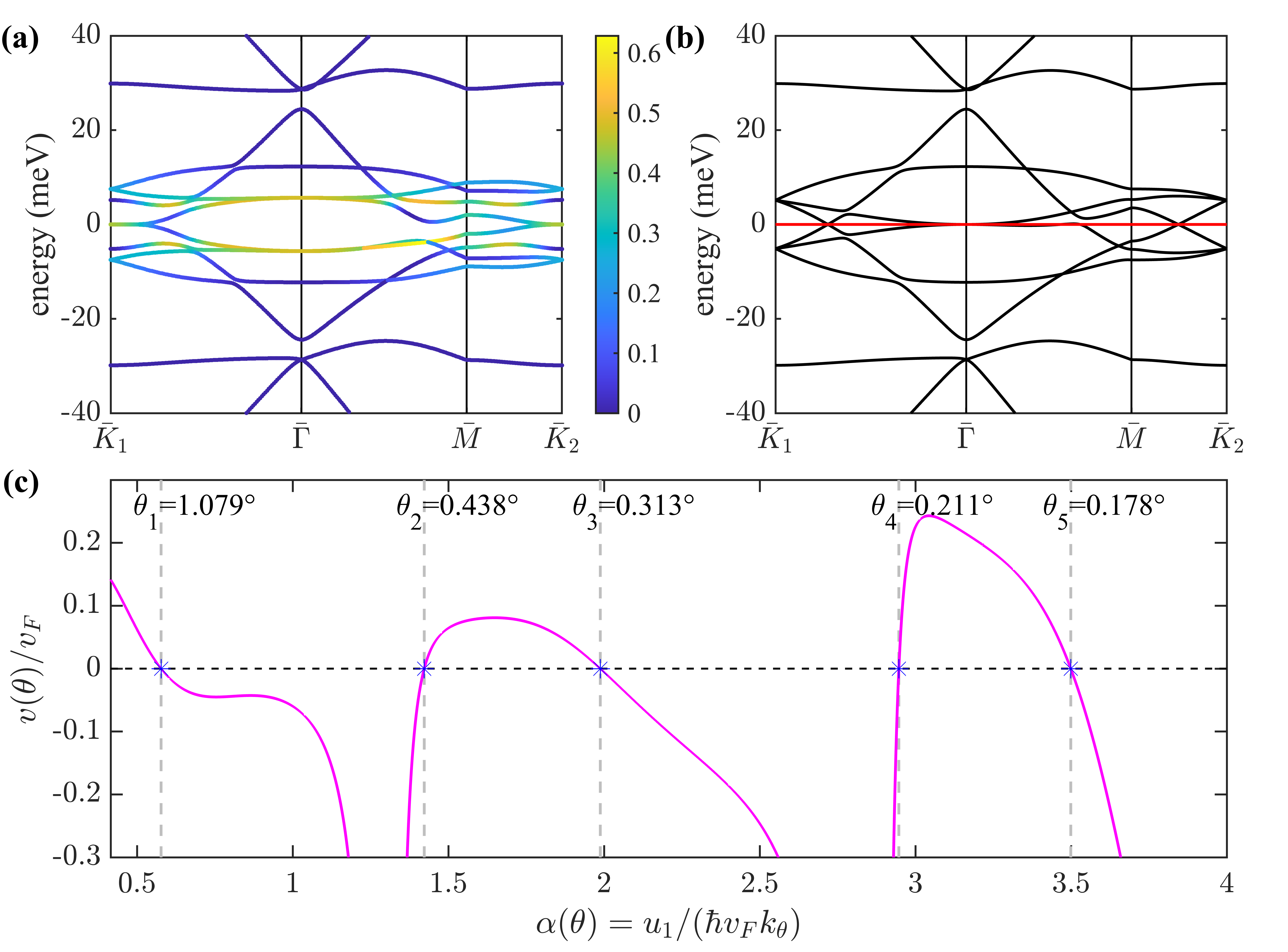}
	\caption{(a) The BM bands and its overlap with ZLLs at the second magic angle $\theta=0.438^{\circ}$. (b) The decoupled bands in the ZLL (red) and OPW (black) subspaces. (c) The Fermi velocity (at $\bar{\tmmathbf{{K}}}_1$) plotted as a function of the dimensionless parameter $\alpha(\theta)$. The first five magic angles (blue stars) are found to be $\alpha_1=0.577$ ($\theta_1=1.079^{\circ}$),
    $\alpha_2=1.422$ ($\theta_2=0.438^{\circ}$), 
    $\alpha_3=1.988$ ($\theta_3=0.313^{\circ}$), 
    $\alpha_4=2.947$ ($\theta_4=0.211^{\circ}$),
    and $\alpha_5=3.499$ ($\theta_5=0.178^{\circ}$).}
    \label{fg:2}
\end{figure}

Fig. \ref{fg:2} shows the BM energy bands and the decoupled bands at the second magic angle $\theta = 0.438^{\circ}$. The lowest two OPW bands become flat around the mBZ center, extending the quadratic touching to a larger range. The complex intersections between the decoupled bands lead to a discrete distribution of ZLL components when the coupling is restored. By keeping the two ZLL bands, only six nearest OPW bands and the couplings among them, an eight-band model can be constructed (not shown), which accurately reproduces the BM bands within the gaps near $\pm25$ meV.

As the angle decreases, more zero-energy PLLs with higher angular momentum can also localize within the moir\'{e} supercell. They are expected to play some roles near the charge neutrality point and might be responsible for the complexity of bands there.

\subsection{Magic angle series: a new perspective}

Before this work, there have been several theoretical attempts to analyze the physical \cite{abelian_field_12prl,pseudo_LL_19prb} or mathematical \cite{chiral_limit_19prl,wkb_21prl} origins of the magic angle series. Our model provides a new perspective to revisit it: without hybridization the ZLL bands are completely flat, whose dispersion will be induced by the coupling to the OPW states. A $k\cdot p$ model focusing on how the ZLLs are altered by OPWs near the moir\'{e} Dirac points can be constructed using the perturbation theory. Such effective Hamiltonian at $\bar{\tmmathbf{k}} \approx \bar{\tmmathbf{K}}_1$ for ZLL orbitals is found to be
\begin{align}
  \mathcal{H}_{\text{eff}}^{\text{zll}} (\bar{\tmmathbf{k}}) \approx
  - \mathcal{H}^{\text{cp}} (\bar{\tmmathbf{k}})
  \frac{1}{\mathcal{H}^{\text{opw}} (\bar{\tmmathbf{k}})}
  \mathcal{H}^{\text{cp} \dagger} (\bar{\tmmathbf{k}}). \label{eq:Heff}
\end{align}
The momentum gradient of $\mathcal{H}_{\tmop{eff}}^{\tmop{zll}}$ at the moir\'{e} Dirac points defines the Fermi velocity operator, which can be expressed using our notations as
\begin{align}
\begin{split}
  \hbar \tmmathbf{v}(\theta) = &- \nabla \mathcal{H}^{\tmop{cp}}_{\bar{\tmmathbf{K}}_1} \frac{1}{\mathcal{H}^{\tmop{opw}}_{\bar{\tmmathbf{K}}_1}} \mathcal{H}^{\tmop{cp}\dag}_{\bar{\tmmathbf{K}}_1}  -\mathcal{H}^{\tmop{cp}}_{\bar{\tmmathbf{K}}_1}
  \frac{1}{\mathcal{H}^{\tmop{opw}}_{\bar{\tmmathbf{K}}_1}} \nabla
  \mathcal{H}^{\tmop{cp\dag}}_{\bar{\tmmathbf{K}}_1} \\
  & +\mathcal{H}^{\tmop{cp}}_{\bar{\tmmathbf{K}}_1}
  \frac{1}{\mathcal{H}^{\tmop{opw}}_{\bar{\tmmathbf{K}}_1}}\nabla
  \mathcal{H}^{\tmop{opw}}_{\bar{\tmmathbf{K}}_1}
  \frac{1}{\mathcal{H}^{\tmop{opw}}_{\bar{\tmmathbf{K}}_1}}
  \mathcal{H}^{\tmop{cp} \dag}_{\bar{\tmmathbf{K}}_1}, 
\end{split}
\label{eq:fermi_vel}
\end{align}
where the subscript $\bar{\tmmathbf{K}}_1$ denotes that all matrices and their gradients are defined at this Dirac point. 

Eq. (\ref{eq:fermi_vel}) makes the direct calculation of the Fermi velocity (no requirement of the eigenenergy) possible. It is found that the Fermi velocity operator always keeps the form: $\tmmathbf{v}(\theta) = -v (\theta) \tmmathbf{\sigma}$. As shown in Fig. \ref{fg:2}(c), the velocity $v (\theta)$ oscillates with $\theta$ and crosses the zero at some discrete angles that are exactly the magic angles. Near some other angles, $v (\theta)$ diverges because the OPW bands touch the ZLL bands (zero energy) there. In this case the above perturbation method fails to predict the actual Fermi velocity. Fortunately, near all first five magic angles, the OPW and ZLL bands are well separated near the Dirac points (see Fig. \ref{fg:2}(b) and SM) so our calculation in these regions is trustworthy. Our study identifies the emergence of magic angles with exactly vanishing Fermi velocity as a typical character of the BM model. In more realistic models like the tight-binding model considering the relaxation effects, this character can be severely smeared. \cite{relaxation4_19prr,21m2D_localized_electron}.

\subsection{Twisted multilayer graphenes}

In this section we discuss the generalization of our model to two types of twisted multilayer graphene (TMG) systems.

The first kind of TMG contains the twisted ($M+N$)-layer graphene systems \cite{TMG1_19prb,TMG2_19prb,TMG3_19prx,TMG7_19nl,TMG5_21sb,TMG6_22prl}. In these materials the twist happens only in the interface of the upper and lower Bernal-stacking multilayers, so we can always treat them as a TBG sandwiched between other graphene sheets. The continuum Hamiltonian can thus be roughly written as (in each valley)
\begin{align}
  H^{\alpha\beta}_{MN} & =  \left(\begin{array}{ccc}
    H_{M - 1}^{\alpha} & T_{\alpha} & \\
    T_{\alpha}^{\dag} & H^{\tmop{BM}} & B_{\beta}\\
    & B_{\beta}^{\dag} & H_{N - 1}^{\beta}
  \end{array}\right), \label{eq:HTMG}
\end{align}
where $H_{M-1}^{\alpha}$ ($H_{N-1}^{\beta}$) is the Hamiltonian of the upper $M - 1$ (lower $N-1$) layers with the stacking chirality $\alpha$ ($\beta$). $T_{\alpha}$ and $B_{\beta}$ represent the tunneling between TBG and its nearest layers. Turning off $T_{\alpha}$ and $B_{\beta}$ gives an isolated TBG subsystem, where the ZLLs and the OPWs can be constructed as usual. In general such systems no longer have strict local zero modes like TBG.

\begin{figure}[ht]
	\centering
	\includegraphics[width=1.0\linewidth]{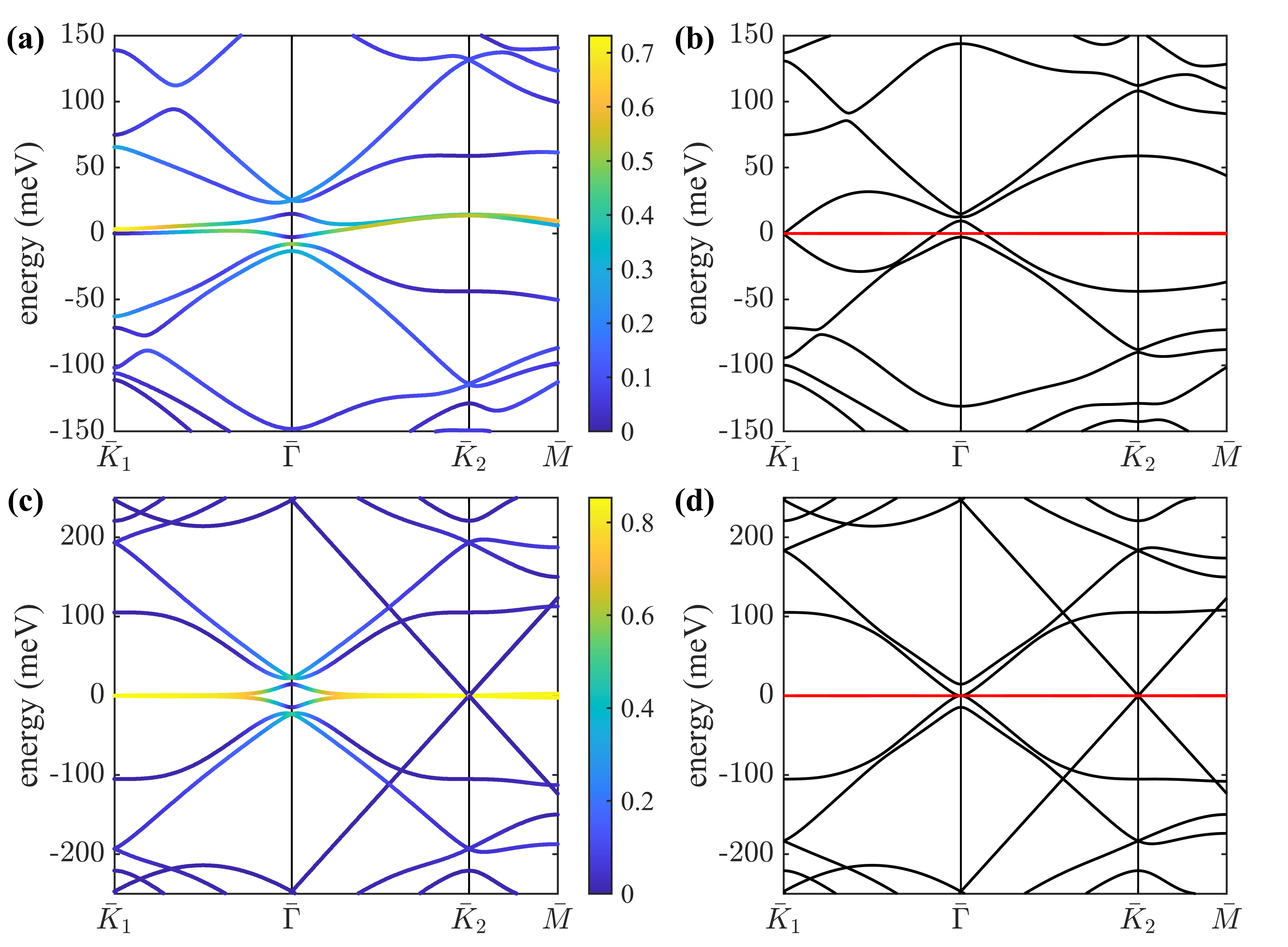}
	\caption{(a) Band structures (the color indicates the overlap with ZLLs) and the decoupled bands in ZLL and OPW subspaces for $1.05^{\circ}$ twisted $1+2$ trilayer graphene. (c) and (d) are the same plots for alternating twisted trilayer graphene with angle $\theta=1.40^{\circ}$. We only show the bands in the valley $\eta=-1$.}
	\label{fg:3}
\end{figure}

The BM bands and the decoupled bands in ZLL and OPW subspaces of the twisted $1+2$ trilayer graphene ($\theta=1.05^{\circ}$) are shown in Fig. \ref{fg:3}(a)(b). One of the common features of twisted ($M+N$) multilayers is the existence of two narrow bands near the charge neutrality point. The ZLLs spread to a larger energy range through the additional inter-layer tunneling, and the OPW bands deviate from the BM bands even away from the mBZ center. The two narrow bands still hold relatively larger ZLL components, in line with the localized states found numerically \cite{TMG7_19nl}. Again at the mBZ center the two flat bands are composed entirely of OPW orbitals.

The second group covers the so-called alternating twisted multilayer graphenes, in which the $l$-th layer is twisted by the angle $(-1)^l\theta/2$. The applicability of our model on such systems relies on the fact that their continuum Hamiltonian can be exactly mapped to a direct sum of some renormalized TBGs (plus a monolayer for odd layers) \cite{19prb_ATMG_by_ashvin,21prb_TTG_superconductivity}, and each of them can be separately treated under our theoretical framework. 

As an example, the trilayer Hamiltonian $H_{\tmop{tri}}$ can be transformed as (in the valley $\eta=-1$)
\begin{align}
  H_{\tmop{tri}}(\tmmathbf{p},\tmmathbf{r}) =  \mathcal{V}\left(\begin{array}{cc}
    H^{\tmop{BM}}(\tmmathbf{p},\tmmathbf{r}) & \\
     & H^{\tmop{D}}(\tmmathbf{p})
  \end{array}\right)\mathcal{V}^{\dag}, \label{eq:ATMG}
\end{align}
where $\mathcal{V}$ is a layer-transformation matrix, $H^{\tmop{D}} = - v_F (\tmmathbf{p} - \hbar\tmmathbf{K}_1)\cdot \tmmathbf{\sigma}$ is the monolayer Dirac cone, and $H^{\tmop{BM}}$ is just the BM Hamiltonian (\ref{eq:BM_Hamiltonian}) with inter-layer couplings replaced by $\sqrt{2}u_0$ and $\sqrt{2}u_1$. Distinguishing the local ZLLs in $H^{\tmop{BM}}$ from all other states leads to the bands shown in Fig. \ref{fg:3}(c)(d). This time both the (effective) monolayer and bilayer host the active itinerant orbitals, while only the latter will couple to the local ZLLs if no strain or external fields exist. It will be interesting to relate our ZLL+OPW representation to the recently proposed heavy-fermion character of this system \cite{21prl_heavy_fermion_TTG}.

\section{Full-band Hartree-Fock calculation} 
\label{sec:4}
\subsection{Formulation of the variational method}
The hybridized ZLL+OPW representation also provides a new viewpoint to revisit correlated insulating states \cite{caoyuan1_18n,exp3_19sci,19n_insulator_SC_LL,20s_QAH_hBN} that have been studied extensively \cite{superconductivity1_18prx,HF1_20prl,HF2_20prb,HF3_20prx,HF4_20prl,HF5_21prb,HF6_21prb,HF7_22prl,HF8_21prr,HF9_22prl}. It was speculated in Ref. \cite{pseudo_LL_19prb} that the Coulomb interaction can split the eight-folded ZLL bands through symmetry breaking, leading to various insulating states. We now quantify this general idea by introducing a new variational method.

The Coulomb interaction is written in plane wave basis as
\begin{align}
  H^{\tmop{I}} = \frac{1}{2 N_{\tmop{m}}} \sum_{\tmmathbf{k} \mu} \sum_{\tmmathbf{k}' \mu'}
  \sum_{\tmmathbf{q}} v_{\tmmathbf{q}} c^{\dag}_{\tmmathbf{k}+\tmmathbf{q},
  \mu} c^{\dag}_{\tmmathbf{k}' -\tmmathbf{q}, \mu'} c_{\tmmathbf{k}' \mu'}
  c_{\tmmathbf{k} \mu},\label{eq:Interaction}
\end{align}
where $\mu \equiv$ ($s$, $\eta$, $\alpha$) is the composite spin, valley and layer/sublattice index. In this work the double-gate screened interaction is adopted,
\begin{align}
 v_{\tmmathbf{q}} =\frac{e^2 \tanh (d_s q)}{2 \varepsilon_0 \varepsilon_s
 \Omega_{\tmop{m}}q}, \label{eq:dg_interaction}
\end{align}
where $q=|\tmmathbf{q}|$, $\Omega_{\tmop{m}} = \sqrt{3} L_{\theta}^2 / 2$ is the area of each moir\'{e} cell, and $\varepsilon_0$ is the permittivity of vacuum. We fix the screening length $d_s = 20$ nm and the dielectric constant $\varepsilon_s = 10$ in this study. 

\begin{figure}[ht]
	\centering
	\includegraphics[width=1.0\linewidth]{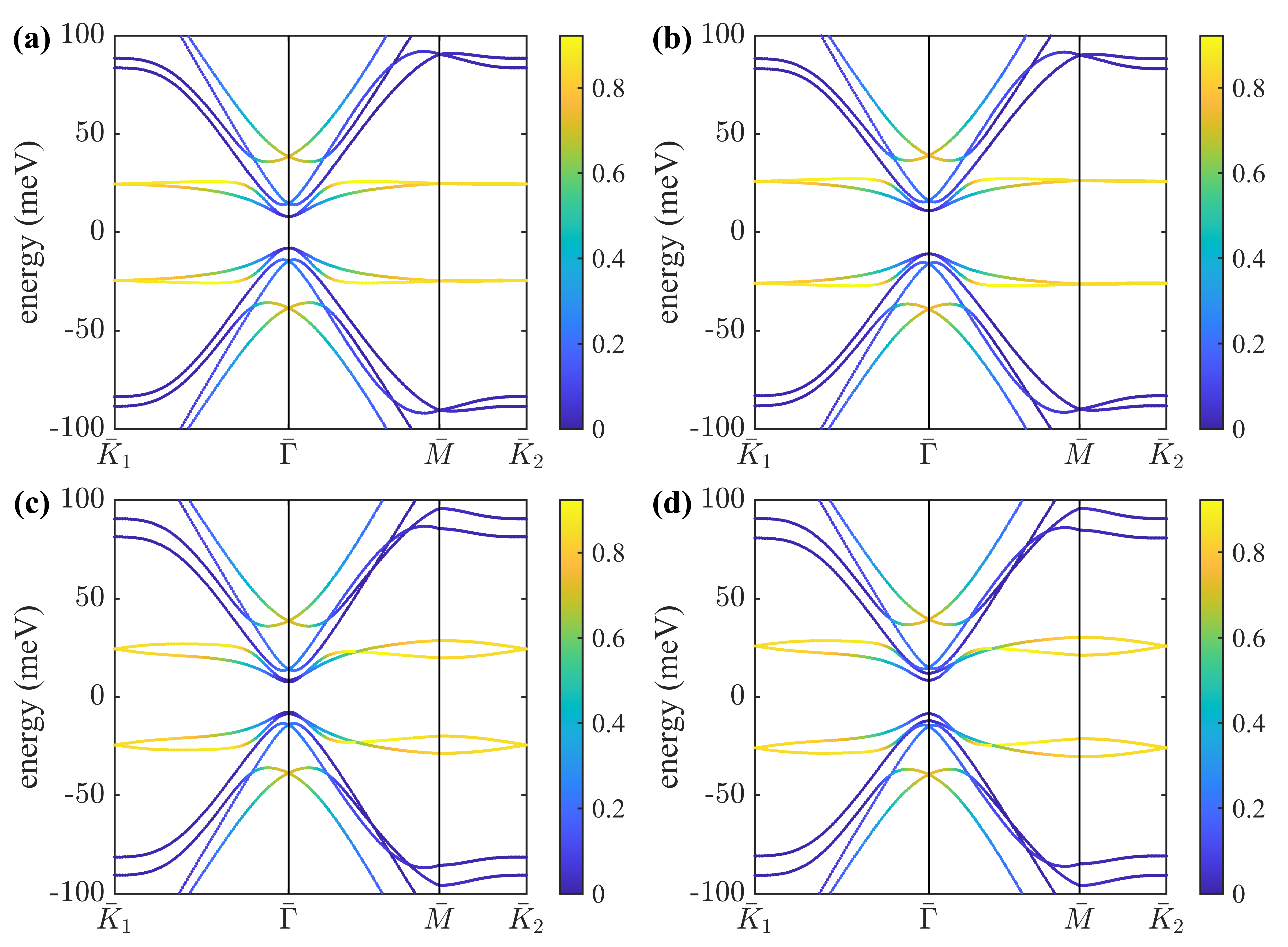}
	\caption{Quasi-particle bands of MATBG at $\nu=0$. (a) and (b) are KIVC bands obtained using the variational method and the self-consistent method, respectively. (c) and (d) are the corresponding VP bands. The color represents the overlap between the $n$-th band with the eight ZLLs: $\sum_{\xi}| \langle\Phi_{\bar{\tmmathbf{k}}, \xi} |\psi_{\bar{\tmmathbf{k}}, n}^{\tmop{HF}}\rangle|^2$, $\xi=(s,\eta,t)$. 
    We have set the chemical potential $\mu_c=0$.}
    \label{fg:4}
\end{figure}

Now let us illustrate our Hartree-Fock (HF) variational method. The trial wave function $|0; \lambda\rangle$ is taken as the ground state of the following ``mean field'' Hamiltonian,
\begin{align}
  H^{\tmop{trial}} = H^{\tmop{BM}} + \sum_l \sum_{\bar{\tmmathbf{k}}}
  \sum_{\xi \xi'} \lambda_l \mathcal{O}^l_{\xi, \xi'}
  f^{\dag}_{\bar{\tmmathbf{k}}, \xi} f_{\bar{\tmmathbf{k}}, \xi'},
  \label{eq:trial_Ham}
\end{align}
where $\xi \equiv$ ($s$, $\eta$, $t$) denotes the composite spin, valley and angular momentum (Chern number) index of ZLLs. $\lambda_l$ ($l = 1$, $2$, $\ldots$, $64$) are variational parameters that take real values. The corresponding order matrices $\mathcal{O}^l$ have the form $s_i \tau_j \sigma_k$ ($i$, $j$, $k = 0$, $x$, $y$, $z$), where $s_i$, $\tau_j$, $\sigma_k$ are Pauli matrices representing the spin, valley and angular momentum degrees of freedom. Given a set of parameters $\lambda_l$, the trial wave function $|0;\lambda \rangle$ offers a single-particle density matrix
\begin{align}
  \rho^{\tmmathbf{G}\mu}_{\tmmathbf{G}'\mu'} (\bar{\tmmathbf{k}}) = \langle 0 ; \lambda | c^{\dag}_{\bar{\tmmathbf{k}}+\tmmathbf{G},\mu} c_{\bar{\tmmathbf{k}}+\tmmathbf{G}',\mu'} | 0 ; \lambda\rangle,
\label{eq:rho}
\end{align}
and the optimal ground state is obtained by minimizing the total energy, which is an implicit function of $\lambda_l$ and can be written through the density matrix as
\begin{align}
e_{\tmop{tot}} = \frac{1}{N_{\tmop{m}}}(E_0[\rho] + E_{\tmop{H}}[\rho] + E_{\tmop{F}}[\rho]).
\label{eq:tot_energy}
\end{align}
Expressions of the kinetic energy $E_0$, Hartree energy $E_{\tmop{H}}$ and Fock energy $E_{\tmop{F}}$ can be found in SM. We adopt a $12\times 12$ $\bar{\tmmathbf{k}}$-mesh sample and 61 $\tmmathbf{G}$ vectors in calculations. Finally, the derivation process of the BM Hamiltonian (\ref{eq:BM_Hamiltonian}) implies that the Hartree-Fock potential at the charge neutrality point has already been included. Therefore, it should be removed from our calculations to avoid double counting \cite{HF2_20prb,HF3_20prx,HF7_22prl,HF8_21prr}. Correspondingly, in our approach the density matrix $\rho$ in the expressions of $E_{\tmop{H}}$ and $E_{\tmop{F}}$ will be replaced by $\tilde{\rho}=\rho - \rho_0$ ($\rho_0$ is the density matrix of the BM Hamiltonian at $\nu=0$). 

\begin{figure}[h]
	\centering
	\includegraphics[width=1.0\linewidth]{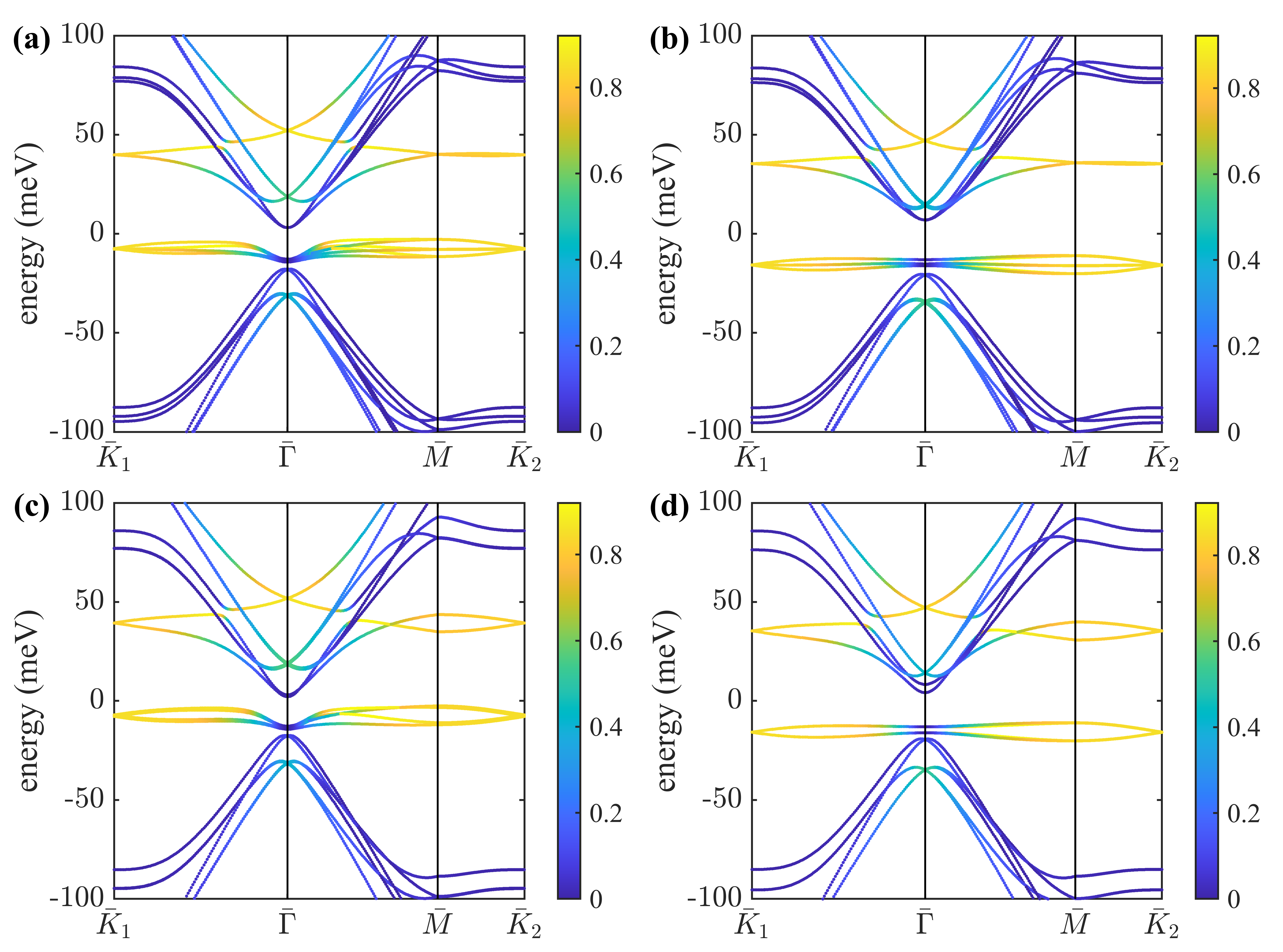}
	\caption{Quasi-particle bands of MATBG at $\nu=2$. (a) and (b) are KIVC bands obtained using the variational method and the self-consistent method, respectively. (c) and (d) show the corresponding VP bands. The color represents the overlap with the ZLL orbitals.
    The chemical potential $\mu_c=0$.}
    \label{fg:5}
\end{figure}

Such variational scheme implies the conjecture that the localized ZLL orbitals play the dominant role in breaking the system into various ordered states \cite{heavy_fermion_22prl,pseudo_LL_19prb}. The itinerant OPWs constitute only $\sim20\%$ of the flat bands. Therefore, we put symmetry-breaking orders only in the ZLL subspace during the variational procedure. More specifically, the second term of Eq. (\ref{eq:trial_Ham}) governs how the eight localized orbitals are combined and splitted by adjusting the order parameters $\lambda_l \mathcal{O}^l$, while the OPWs and the couplings between them are left unchangeable.

\subsection{Ground states at integer fillings}

In this subsection we present the main results for MATBG at integer fillings $\nu=0$, $1$, $2$, $3$. Minimizing the total energy (\ref{eq:tot_energy}) generates several gapped insulating states at each filling. The numerical results are summarized in the Table \ref{tb:ground_states} (other ordered states with higher energies are not shown).

\begin{table*}[]
\caption{The order parameters $\mathcal{O}^l$, order strength $\lambda_l$, and the condensation energy obtained through the variational method $e_{\tmop{Var}}$ and self-consistent method $e_{\tmop{SCMF }}$ for the KIVC and VP states at integer fillings. All numbers have the units meV.}
\label{tb:ground_states}
\centering
\begin{tabular}{c|c|c|c|c|c}
\hline\hline
$\nu$ & order & $\mathcal{O}^{l}$   & $\lambda_l$ & $e_{\tmop{Var}}$ & $e_{\tmop{SCMF}}$ \\ \hline
\multirow{2}{*}{0}&KIVC&$\tau_x\sigma_z$ &$25.233$ &$-50.658$ &$-56.135$ \\ 
~ &VP &$\tau_z$ &$21.565$ &$-51.044$ &$-54.661$ \\ \hline
\multirow{2}{*}{1}&KIVC&$I_0,\sigma_z,\tau_x,\tau_x\sigma_z,s_z,s_z\sigma_z,s_z\tau_x,s_z\tau_x\sigma_z$ &$14.265,7.245,6.814,19.465,5.653,-7.263,-6.839,6.535$ &$-48.371$ &$-54.860$ \\
~&VP &$I_0,\sigma_z,\tau_z,\tau_z\sigma_z,s_z,s_z\sigma_z,s_z\tau_z,s_z\tau_z\sigma_z$ &$13.711,7.152,16.938,7.175,5.179,-7.174,5.179,-7.174$ &$-48.581$ &$-54.083$  \\ \hline

\multirow{2}{*}{2}&KIVC&$I_0,s_z,\tau_x\sigma_z,s_z\tau_x\sigma_z$ &$25.758,12.611,14.038,14.038$ &$-43.866$&$-54.923$ \\
~&VP&$I_0,s_z,\tau_z,s_z\tau_z$ &$24.511,11.540,12.052,11.538$&$-43.727$&$-54.196$  \\ \hline

\multirow{2}{*}{3}&KIVC&$I_0,\sigma_z,\tau_x,\tau_x\sigma_z,s_z,s_z\sigma_z,s_z\tau_x,s_z\tau_x\sigma_z$ &$33.486,7.971,8.024,7.194,7.823,7.971,8.024,7.194$&$-35.782$&$-52.755$ \\
~&VP&$I_0,\sigma_z,\tau_z,\tau_z\sigma_z,s_z,s_z\sigma_z,s_z\tau_z,s_z\tau_z\sigma_z$&$33.458,7.957,6.984,7.989,6.991,7.951,7.704,7.956$&$-35.707$&$-52.719$  \\ \hline\hline

\end{tabular}
\end{table*}

At $\nu = 0$, the convergence can be well obtained by introducing only one order parameter for each ordered state. Two competitive groups of states are found to have lower energies than others. The first group is the Kramers inter-valley coherent states (KIVC) \cite{superconductivity1_18prx,HF3_20prx} with the condensation energy $-50.658$ meV, whose order parameter can be fixed as $\tau_x \sigma_z$. The second group includes the valley-polarized state (VP: $\tau_z$), spin-polarized state (SP: $s_z$) and spin-valley-locked state (SVL: $\tau_z s_z$). All these flavor-polarized states have exactly the same energy $-51.044$ meV. The quasi-particle bands for KIVC state and VP state are shown in Fig. \ref{fg:4}. Since they have very close energies, we may just treat the KIVC states and flavor-polarized states as two degenerate candidates of the actual ground state.

At other fillings, some flavor degeneracies of the above low-energy states will be further lifted, and more variational parameters are involved to split the eight ZLL orbitals. Take the KIVC state as an example, four dominant order parameters $I_0=s_0 \tau_0 \sigma_0$, $s_z$, $\tau_x \sigma_z$, $s_z \tau_x \sigma_z$ are necessary at $\nu = 2$, and other four orders $\sigma_z$, $s_z \sigma_z$, $\tau_x$, $s_z \tau_x$ are also essential at $\nu = 1$, $3$ to remove the spin degeneracy. The quasi-particle bands at $\nu=2$ are shown in Fig. \ref{fg:5}.

In our HF variational approach the symmetry breaking order parameters are limited within the ZLL subspace, which is completely local and $\bar{\tmmathbf{k}}$-independent. Therefore, the number of the variational parameters is at most 12 in our approach (if we fix the gauge), which makes the calculation much feasible. Once the convergence is obtained, the resulting ground state then provides a starting state to perform the full self-consistent mean-field calculation, where the variational parameter will be the completely $\bar{\tmmathbf{k}}$-dependent single-particle density matrix introduced in Eq. (\ref{eq:rho}).  As shown in Figs. \ref{fg:4} and \ref{fg:5}, the HF bands obtained using these two methods are quite similar, which strongly supports our conclusion that the correlation effects in MATBG are mainly limited within the ZLLs, rather than the OPWs. 

Finally we note that the strong coupling conjecture here may collapse in the chiral limit $u_0=0$, and a brief discussion is given in SM [URL will be inserted by publisher].

\section{Summary} 
\label{sec:5}
In conclusion, we have proposed a new representation for TBG which clearly distinguishes the local ZLL orbitals from all other itinerant OPW states. The ZLLs are the exact zero-mode eigenstates of the BM Hamiltonian near the AA-stacking center. They have similar properties to the Wannier functions given in Ref. \cite{heavy_fermion_22prl}. Besides, they also have a clearer interpretation as the generalized zeroth Landau levels of Dirac fermions and can be analytically determined from the model parameters. The BM flat bands near the magic angle are then understood as the outcome of the interplay between the local ZLLs and the itinerant OPWs. Due to the universal existence of ZLLs, this model can be applied to smaller-angle TBG and TMG systems. A Fermi velocity with a sign can be defined and calculated for TBG with the help of this representation, which successfully explains the robustness of the magic angle series.

The clear division of localized and itinerant components in the band structure of TBG provides a feaible way to treat the correlation and topological features of TBG at the same time. As we have demonstrated in the present study, the correlation effects only need to be considered within the ZLLs rather than the OPWs, which greatly mimics the situation of heavy fermion materials. Next, it will be interesting to look at various of correlation effects generated by the coupling between the ZLLs and OPWs especially for the non-integer doping, including the Kondo physics, some possible heavy fermion behaviors, the RKKY-coupling-induced symmetry breaking order, and the superconductivity.

\textit{Note added.} Recently a new superconducting theory in TBG appeared \cite{sup_heavy_arxiv}, which is based on a similar picture of local electrons hybridized with the itinerant ones. In that study the pairing attraction is assumed to involve exclusively the local orbitals.

\section*{Acknowledgements} 
\label{sec:acknowledgements}
We thank  Professor Zhida Song, Jianpeng Liu and Andrei Bernevig for helpful discussions. X. D. acknowledges financial support from the Hong Kong Research Grants Council (Project No. GRF16300918 and No. 16309020).

\bibliography{refs} 

\end{document}